# Emerging electro-optic molecular crystals for optoelectronic integration

*Keishi Sunami[1]\*, Sachio Horiuchi[1], Yoriko Sonoda[1], Naomi Fujiki[1], Toshiki Higashino[1], Yuki Atsumi[2], Shoji Ishibashi[1], and Jun'ya Tsutsumi[1]*

[1]*Core Electronics Technology Research Institute (CETRI), National Institute of Advanced Industrial Science and Technology (AIST), Tsukuba, Ibaraki 305-8565 Japan*

[2]*Photonics-Electronics Integration Research Center (PEIRC), National Institute of Advanced Industrial Science and Technology (AIST), Tsukuba, Ibaraki 305-8568 Japan*

\*Corresponding author: k.sunami@aist.go.jp (K.S.), junya.tsutsumi@aist.go.jp (J.T.)

**Abstract**

The rapid advancement of communication technology necessitates the development of hybrid optical modulators that integrate silicon photonics with electro-optic (EO) materials to enable ultrafast, low-power, and compact photonic devices. However, no existing material simultaneously meets the requirements for high EO performance, process compatibility, and thermal stability, which are essential for practical optoelectronic integration. Here, we present 4-(4'-nitrophenylazo)diphenylamine (NDPA) and its derivative, novel nonlinear optical molecular crystals discovered through materials screening incorporating crystal habit prediction. They demonstrate crystallization into high-quality aligned thin films and ultrafine silicon slot fillings through the capillary action of melts. The resulting EO performance much exceed that of conventional lithium niobate and are comparable to state-of-the-art EO polymers, with excellent thermal stability maintained for over 1000 hours at 393 K. The present EO materials satisfying all the requirements above are promising candidates for silicon-organic hybrid optical modulators, opening a significant step toward scalable and high-performance optoelectronic technologies.

## 1 Introduction

A massive increase of connected devices and a rapid advancement of artificial intelligence cause a significant rise in data traffic, making high-speed and energy-efficient data communication an urgent challenge. The promising solution is optoelectronic integration to combine optical communication with electronic data processing within a silicon chip or a silicon package.[1,2] This concept relies on silicon photonics technologies which allow fabricating optical circuits on silicon chips.[3–5] Among the technologies, an on-chip optical modulator plays a vital role on electrical-to-optical signal conversion. Since conventional on-chip optical modulators, exploiting the free-carrier plasma dispersion effect of silicon,[6,7] are approaching their performance limit, there is a strong demand to transition toward modulators based on the high-speed and energy-efficient electro-optic (EO) effect. However, silicon lacks the Pockels effect, i.e. EO-inactive, so that considerable research efforts have been directed to integrate EO-active materials onto silicon waveguides.[8,9]

Requirements of EO materials toward the optoelectronic integration are not only excellent EO activity but also CMOS-process compatibility and high thermal stability.[2] Excellent EO activity is essential to miniaturize optical modulators toward scalable photonic integrated circuits. CMOS-process compatibility is critical for mass production in semiconductor foundries, while high thermal stability is indispensable for integrating optical modulators with heat-generating electronic circuits. One of the promising candidates is lithium niobate. It is a well-established EO material with low optical loss and high thermal stability and can be integrated into silicon through transferring and low-temperature bonding of lithium niobate on insulator (LNOI).[8,10] Nevertheless, several challenges remain, including limited process yield due to alignment tolerances and bonding interface quality, as well as the overall complexity of the integration process. Another candidate is an EO polymer which recently attracts attention as a new class of EO materials.[11–15] It has EO activity much higher than the lithium niobate and can be easily deposited on silicon waveguide through solution process. Furthermore, by combining with a slot waveguide structure,[16] it allows highly efficient optical modulation, enabling significant miniaturization of optical modulators. However, it requires high electric-field poling to align chromophores and inherently suffers from thermal instability due to orientational relaxation of the aligned chromophores. To improve thermal instability, there have been efforts such as elevating glass transition temperature and crosslinking polymer backbone,[17,18] whereas they introduce trade-offs: The former reduces poling efficiency, while the latter degrades optical transparency due to increased light scattering.

EO molecular crystals, exemplified by 4-dimethylamino-*N*-methyl-4-stilbazolium tosylate (DAST)[19–21] and 2-(3-(4-hydroxystyryl)-5,5-dimethylcyclohex-2-enylidene) malononitrile (OH1),[22,23] exhibit nonlinear optical activity as high as EO polymers, so that they have been utilized especially for terahertz photonics applications.[23] As they are crystalline materials whose chromophores spontaneously align without poling, they are inherently free from thermal instability caused by orientational relaxation in sharp contrast to EO polymers. Despite these advantages, their application has been limited to discrete optical components. The main reason not to be applied to photonic integrated circuits is poor film formability due to their anisotropic crystal growth although there are a few exceptions such as *N*-benzyl-2-methyl-4-nitroaniline (BNA).[24]

Here we report novel EO molecular crystals that excel in all of EO activity, film formability, and thermal stability. Systematic materials screening was conducted on the Cambridge structural database (CSD) with introducing parameters such as hyperpolarizability, crystal habit, and melting point which are relevant to EO activity, film formability, and thermal stability, respectively. Then, 8 kinds of extracted EO molecular crystals were applied for melt-casting experiments to examine their film formability. Consequently, two materials were found to exhibit excellent film formability due to two-dimensional crystal growth: 4-((4-nitrobenzylidene)amino)-*N*-phenylaniline (PNPA) and 4-(4'-nitrophenylazo)diphenylamine (NDPA). Remarkably, their EO activities were comparable to state-of-the-art EO polymers and were stable over 1000 h under high temperature condition of 393 K. These properties highlight their potential as next-generation EO materials for integrated photonics.

## 2 Results and Discussion
### 2.1 Materials screening by crystal habit prediction

Conventional materials screening for EO materials has employed only two parameters with no exception, i.e. molecular hyperpolarizability $\beta$ and molecular orientation within a crystal. This is reasonable to find high EO-performance materials because highly oriented chromophores with large hyperpolarizability result in superior EO activity. For example, molecular design for enlarging $\beta$ is to attach strong electron donating and accepting substituents at both terminal of extended π-conjugation backbone,[25,26] as exemplified by the materials displayed in Figs. 2 and S1 and Table 1. However, such a screening alone is not sufficient for high film formability and thermal stability that are required toward optoelectronic integration of the EO materials. Therefore, the materials screening in this study introduced a crystal-habit parameter $I_{2D}$ and a melting point $T_m$ which are indicators for the film formability and the thermal stability, respectively. $I_{2D}$ was obtained as follows: Crystal habit was simulated for various molecular crystals recorded in CSD by the surface attachment energy calculation program, HABIT,[27,28] with the Gavezzotti's empirical force field.[29,30] Then, its anisotropy estimated by ellipsoid fitting was plotted in a triangle graph as a function of $b/a$ and $c/a$ where each corner corresponds to 1-dimensional crystal growth ($b/a = c/a = 0$), 2-dimensional crystal growth ($b/a = 1, c/a = 0$), and 3-dimensional crystal growth ($b/a = c/a = 1$), respectively (Fig. 1a). $I_{2D}$ was defined as the distance from the 2-dimensional crystal growth, so that high film formability is expected for small $I_{2D}$. Fig. 1b shows $T_m$ of EO molecular crystals plotted as a function of $I_{2D}$ with the $\beta$ value (see Figs. 2 and S1 for the molecular structures). Each point corresponds to EO molecular crystals extracted from CSD that meet the following criteria: They consist of single component D-π-A chromophores where electron-donating (D) and electron-accepting (A) functional terminal groups are connected by π-conjugated bonds. Their total hyperpolarizability (non-resonant) $\beta$ simulated by a quantum chemical calculation are as high as $50 \times 10^{-40} \sim 520 \times 10^{-40}$ m$^4$ V$^{-1}$. Their angle $\theta$ between molecular hyperpolarizability vector and crystal polar axis ranges 0~30° so that their EO coefficients (second order NLO susceptibility) are nearly maximized along the crystal polar axis. Note that the diagonal tensor component of EO coefficient is proportional to cos$^3\theta$, meaning large EO coefficient for small $\theta$. Among the materials plotted in Fig. 1b, the ones with small $I_{2D}$ are favorable in terms of film formability. On the other hand, regarding the melting point, the moderate $T_m$ ranging approximately from 420 to 480 K is favorable as indicated by the shaded zone in Fig, 1b, because the lower $T_m$ causes thermal instability and the higher $T_m$ reduces the processability, especially for melt casting. Thus, NDPA, PNPA, OH1, and (*E*)-3-(4-carbazol-9-ylphenyl)-1-(4-nitrophenyl)prop-2-en-1-one (CPNC), which satisfy the criteria of $\beta > 200 \times 10^{-40}$ m$^4$ V$^{-1}$, $I_{2D} < 0.6$, and $T_m \sim 450$ K, were extracted as the EO molecular crystals with high EO activity, film formability, and thermal stability (see Fig. 2 and Table 1).

**2.2 Film-formability test by melt casting**

Film formability of the EO molecular crystals extracted via materials screening were experimentally examined by melt casting. For comparison, the experiments were also conducted for BNA, 4-*N*,*N*-dimethylamino-4'-cyanobiphenyl (DMACB), 1-(4-nitrophenyl)-1*H*-pyrrole (NPRO), and 2-methyl-4-nitroaniline (MNA) which do not satisfy one or more of the screening criteria for $I_{2D}$, $\beta$, and $T_m$ mentioned in Sec. 2.1. As shown in Fig. 3a, the powder samples sandwiched by the upper and lower substrates were heated above $T_m$ so as to melt and spread along the gap between the two substrates due to capillary action. After cooling slowly, they solidified to form thin films. In this study, solution-based processes such as spin coating were not employed because they often cause inhomogeneity in crystal orientation due to non-uniform solvent evaporation, resulting in the formation of small crystallites and polycrystalline domains. Fig. 3b shows the result of melt casting. The crystallite size increased as the order of NPRO (< 50 μm) < MNA (< 200 μm) < DMACB (> 1000 μm) < PNPA (> 1000 μm ) < NDPA (> 1000 μm), which roughly obeyed the decrease of $I_{2D}$: MNA (0.67) > NPRO (0.66) > DMACB (0.59) > PNPA (0.53) > NDPA (0.51). However, it decreased in CPNC (< 50 μm), OH1(< 1 μm), and BNA (< 50 μm) despite small $I_{2D}$: CPNC (0.51), OH1 (0.43), and BNA (0.37). This deviation can be ascribed to super cooling. As shown in Table 1 and Figs. S2c,d,h, these materials exhibited noticeable supercooling ($T_m - T_s > 100$ K), which interfered crystal growth due to abrupt crystallization. It should be noted that DMACB solidified to an EO-inactive polymorph through the melt casting, as confirmed by DSC and SHG experiments (see Supporting Information and Fig. S2e). The remainder PNPA and NDPA, whose melting points and crystal habit predictions are in the preferable range, do not suffer from neither undesired supercooling nor polymorphism.

Consequently, these two EO molecular crystals afforded to form single crystal films with the size of > 1000 μm easily. Note that they are reported to be isostructural[31–33] and gave almost isomorphous film morphologies; these films were confirmed to possess the in-plane polar axis, which is parallel to the [102] direction within the (010) plane (Supporting Information). This feature is suitable for the application of EO modulator, especially in the slot waveguide configuration.

For PNPA and NDPA, to demonstrate the potential for integration with optoelectronic device, the fabrication of thin films within ultra-narrow silicon slots mimicking slot waveguides by the melt casting was examined. We prepared the silicon substrate with rectangular slots, the shapes of which range from 140 nm to 3 μm in width, 200-300 nm in depth, and 15 mm in length by lithography (see Fig. 3c). The powder samples were loaded into the pool at the end of the slots, and then the substrates were heated so that the EO melts flow into the slots via capillary action. Figures 3b shows the crossed-Nicols images of the resulting films within the silicon slots. For the slots with the width of ≥190 nm, they appear to be homogeneously colored due to the birefringence of EO materials, clearly indicating the successful filling of not polycrystalline but single-crystalline PNPA films. In a similar way, NDPA was also filled in the 140 nm-wide slot (Supporting Information). Given that the slot waveguides typically have 100–200 nm gaps, highly ordered uniform PNPA and NDPA films can be easily integrated into such device structures.

## 2.3 EO measurements

Let us uncover the EO performance of PNPA and NDPA thin films. Recently, the bulk crystal of PNPA was reported to show highly efficient terahertz wave generation.[31,34] Therefore, it is expected that PNPA and its isostructural material NDPA potentially demonstrate excellent EO performance in the thin film crystal forms as well. We employed the birefringence ferroelectric field-modulation imaging (FFMI)[35] for the EO evaluation. Applying the electric field to EO materials, the change in the refractive indexes through the EO effect yields the slight modulation of the birefringence of EO materials. This technique detects it as the change in the transmitted light intensity $I_\perp$, $\Delta I_\perp$, by using a cross-polarized optical microscope (see Section 4 for details). We used a complementary metal–oxide–semiconductor (CMOS) image sensor as a detector (see Fig. S6), enabling spatially resolved measurements of EO performance over thin films.

To apply a transverse electric field to the EO films, interdigitated gold electrodes (Line/Space = 5/5 μm; DropSens, Spain) deposited on a glass substrate were used. We selected the domain whose polar axis is parallel to the electric field direction (surrounded by the dotted line in Fig. 4a) for measuring the EO performance. Figures 4b and 4c show the crossed-Nicols ($I_\perp$) and modulation images ($\Delta I_\perp/I_\perp$) for PNPA obtained by FFMI at the wavelength of the incident light $\lambda$ = 810 nm. In the relevant region exhibiting the bright crossed-Nicols image due to the birefringence, the modulation signal was also clearly observed. This is reasonable because EO effect is maximized when the polar axis is parallel to the electric field. Note that the red and blue colors alternate in adjacent lines in Fig. 4c, meaning that the sign of $\Delta I_\perp/I_\perp$ is reversed (Fig. 4e) because of the electric field in the opposite direction on either side of the interdigitated electrodes (Fig. 4a). The modulation signal induced by the EO effect is much affected by the phase retardation of elliptically polarized outgoing light. In our measurements, we accurately controlled the phase retardation using the variable retarder set after the sample and measured $\Delta I_\perp/I_\perp$ under the various phase retardation values (see Fig. 4d and Section 4). From these measurements, the $n_3^3 r_{33}$ values were evaluated to be ~1050 (1250) pm/V for NDPA and ~700 (1200) pm/V for PNPA at $\lambda$ = 810 (730) nm (Table 2). Here, we set the directions parallel and perpendicular to the polar axis within the (010) plane (that is, the [102] and [-20-1] directions, respectively) to the EO tensor indices 3 and 1. In these evaluations, other terms such as $r_{13}$ and $r_{53}$ are neglected because they are expected to be negligibly small relative to $r_{33}$ in the highly anisotropic crystal structure of NDPA and PNPA where the molecular hyper polarizability vectors align almost the same crystal direction. The enhancement of $n_3^3 r_{33}$ at shorter wavelengths is due to the resonance effect near the absorption wavelength (corresponding to the intramolecular transition energy). The obtained EO performance of PNPA and NDPA significantly exceed that of LN ($n_3^3 r_{33}$ ~ 320 pm/V) and comparable to that of EO polymers in device ($n^3 r_{33}$ ~ 700 pm/V).[36] It is noteworthy that these materials not only demonstrate high processability for thin film fabrication but also achieve excellent EO performance.

**2.4 Thermal stability test**

We tested the thermal and long-term stabilities of EO performance for PNPA. Through the measurements at high temperatures up to 393 K, the $n_3^3 r_{33}$ values of PNPA were confirmed to be temperature independent (Fig. 5a). The DSC analysis also indicated that PNPA exhibits only a peak by melting without any decomposition or phase transitions below $T_m$ = 452 K (the inset of Fig. 5a). Moreover, we conducted the accelerated aging test at 393 K in a nitrogen atmosphere. In Fig. 5b, we plotted the root mean square values of the modulation signals $\Delta I_\perp/I_\perp$, evaluated over the area of >250 × 250 μm$^2$ in the PNPA film fabricated on the interdigitated electrode substrate, both with and without sealing using a glass plate (Supporting Information). Without sealing, $\Delta I_\perp/I_\perp$ degraded by 15% after 200 hours (Fig. 5b). This degradation was observed not only in the modulation image but also as a visible change in the optical microscope image (Supporting Information). In contrast, when sealed with a glass plate, $\Delta I_\perp/I_\perp$ showed almost no degradation (within 5%) even after 1000 hours. The strong suppression of degradation by sealing suggests that it is caused by local sublimation of the film surface rather than thermal decomposition. This outstanding thermal stability completely surpasses that of EO polymers, in which the EO performance degradation of 20-30% occurs at 393 K due to orientational relaxation of the aligned chromophores,[37] highlighting the significant advantage of crystalline EO materials. Compared to the widely accepted Telcordia GR-468-CORE standard (85 °C [= 358 K]/2000 hours) for optical device reliability,[38] the present evaluation was conducted at a higher temperature of 393 K, under which the EO performance remained stable for over 1000 hours. The ability to retain EO activity under such stringent conditions underscores the exceptional thermal stability of PNPA; notably, this material uniquely combines high EO performance, film formability, and long-term thermal stability, which is rarely achieved simultaneously in EO materials.

**2.5 Spontaneous ordering of crystal orientation by temperature gradient**

To integrate an EO molecular crystal to an optical modulator with maximized EO efficiency, their crystal polar axis should be controlled to be parallel to applied electric field. On the other hand, the film orientation by the melt casting mentioned in Section 2.3 was random. Here, we demonstrate spontaneous ordering of crystal orientation by temperature gradient. A temperature gradient was generated between both ends of the substrate, with the low-temperature side set below $T_m$ and the high-temperature side above $T_m$. By lowering the temperature of the high-temperature side slowly, the thin film growth can be controlled to proceed along the temperature gradient where the nonmelted portion at the low-temperature side works as a seed crystal. Figure 6a shows the crossed-Nicols micrograph of PNPA thin films fabricated under the temperature gradient using two different cooling rates for the high-temperature side: 30 K/min (rapid cooling) at the upper half of the domains and 3 K/min (slow cooling) at the lower half of the domains. A clearly oriented film along the temperature gradient was successfully obtained in the slow-cooled region. We show a histogram representing the distribution of domain orientation angles relative to the temperature gradient direction (Fig. 6b). It was revealed that the film in the rapid-cooled region exhibited random orientation, whereas the slow-cooled region showed distinct peaks in the domain distribution around 30° and -60°. These peak angles correspond to the tilt of the polar axis (∥ [102]) with respect to the film growth direction (∥ [001]) in PNPA (Supporting Information). These results indicate that the desired crystal orientation for integration into optical modulators can be achieved by controlling both the direction and cooling rate of temperature gradient.

Compared to electric-field poling, the temperature-gradient orientation control offers a significant advantage for fabricating large-area oriented films. Even under the preliminary experimental condition, the longitudinal size of the domain with controlled alignment exceeds 3 mm as shown in Fig. 6a. In the case of fabricating a slot-waveguide modulator using PNPA, the device length required for π phase shift ($V_\pi L$) was estimated to be ~0.5 mm, based on the equation $V_\pi L = \lambda g/2n_3^3 r_{33}\Gamma$,[39] which is well within the aligned domain fabricated in this study. This estimation assumed $n_3^3 r_{33}$ = 700 pm/V, $\lambda$ = 810 nm, the applied voltage of 1 V, the electrode gap $g$ = 190 nm (corresponding to the narrowest slot width successfully filled with PNPA in this study (see Fig. 3c)), and the interaction factor between the electrical and optical electric field $\Gamma$ = 0.2, a typical value in silicon-organic hybrid modulators.[39] Further optimization of the temperature gradient and cooling conditions is expected to enable orientation control over larger areas, suggesting the high potential of this approach for large-scale integration in optoelectronic devices.

**2.6 Material design of EO molecular crystals for optoelectronic integration**

Finally, we discuss the exceptional performance of EO activity, thin-film formability, and thermal stability observed in PNPA and NDPA from a material design perspective. Both molecules are composed of strong donor and acceptor groups connected via an extended π-conjugated bond, which is responsible for the large calculated hyperpolarizability $β$ (Table 1). Such large π-conjugation typically leads to red-shifted absorption edges due to the reduction of the HOMO-LUMO gap; a clear trend can be observed among NDPA, PNPA, OH1 and CPNC with relatively similar molecular structures as seen in the $λ_{th}$ values summarized in Table 1. The red shift of the absorption edge yields enhanced NLO activity due to the resonance effect close to the measurement wavelength. In addition, the crystal EO performance of PNPA and NDPA is significantly enhanced by the perfectly aligned molecular orientation within the crystalline lattice. The relative orientation between molecular hyperpolarizability vector and crystal polar axis shows $\cos^3 θ \sim 0.9998$ in these materials,[40] which is ideal for translating microscopic $β$ into macroscopic EO coefficients. Regarding thermal stability, the extended π-conjugation also tends to correlate with the increased $T_m$, leading to their robust EO performance at elevated temperatures. Thin-film formability, in contrast, is not trivially linked to molecular structure. In this study, we adopted a crystal habit-based approach and selected the materials with 2D growth. Consequently, we successfully demonstrated high-quality film formation in PNPA and NDPA, suggesting the effectiveness of this approach. Thus, the combination of the molecular design for high EO response and the crystal habit prediction for 2D growth presents a promising pathway for the exploration of EO molecular crystals suitable for optoelectronic integration.

**3  Conclusion**

In summary, we explored the novel EO molecular crystals, NDPA and PNPA, that exhibit high EO performance with $n_3^3 r_{33}$ values exceeding 700 pm/V, surpassing that of conventional material of LN and being comparable to existing EO polymers. These materials also demonstrate excellent film formability and thermal stability up to 393 K, fulfilling the critical requirements for optoelectronic integration. The orientation control of thin film by temperature-gradient melt casting offers a scalable approach to achieving alignment without electric-field poling, enabling the convenient fabrication of uniformly aligned films over large areas. The excellent integration ability into narrow slots (e.g., 140–190 nm) further highlights the practical advantages of these materials for integrated slot waveguide devices. The present discovery was guided by materials screening strategy based on crystal habit prediction, where we focused on identifying molecules with high 2D crystal growth. Among them, we selected candidates with large $β$ and high $T_m$, enabling not only efficient EO response but also robust device operation under elevated temperatures. Historically, molecular crystals have been largely overlooked in the context of silicon photonics, mainly due to concerns about poor film formation. However, the successful fabrication of thin-films with high and thermally stable EO performance in PNPA and NDPA demonstrates the applicability of molecular crystals in silicon-organic hybrid photonic devices. This breakthrough suggests the potential to outperform polymer-based EO devices, especially in large-area applications under high thermal conditions. Thus, the present findings address current limitations of existing EO materials and establish molecular crystals as compelling candidates for next-generation large-scale photonic technologies, including a high-speed optical modulator, an optical phased array, and a spatial light modulator.

## 4 Experimental Section

**General**

PNPA and CPNC were prepared according to the reported procedure,[31,41] and NDPA (known also as the Disperse Orange 1 (DO1)) was purchased from Angene Co. Ltd. They were purified by repeating twice the vacuum sublimation under temperature gradient. The prismatic crystals of PNPA with well developed (1-10) planes were grown by slow evaporation of dichloromethane solution, whereas the brownish red elongated plates of NDPA crystals developed with the (010) planes were obtained during the sublimation process.

The other commercially available OH1 (TakachihoSangyo), DMACB (ChemCollect GmbH), BNA (Angene Co. Ltd.), MNA and NPRO (Tokyo Chemical Industry) were purified by the vacuum sublimation under temperature gradient and recrystallization before use.

**Diffuse reflectance spectra measurements**

Diffuse reflectance spectra were measured using a Shimadzu UV-3150 spectrometer equipped with an integrating sphere accessory (Shimadzu ISR-3100) and transformed into absorption spectra by Kubelka−Munk function. The absorption threshold wavelength $λ_{th}$ was determined by linearly extrapolating the steep falloff region of the spectra on the long-wavelength side to zero intensity.

**DSC measurements**

The thermal analysis was performed using a differential scanning calorimeter (DSC7000X; Hitachi High-Technologies Corp., Tokyo). The sample was encapsulated in an aluminum pan and heated at a rate of 5 K/min. The temperature was calibrated by using the melting point of indium (429.8 K). The melting points determined by DSC agree with the reported maxima[31,42–47] within 3 K except for the CPNC without the reported data.

**Calculation of hyperpolarizability**

We calculated the hyperpolarizability $\beta$ by Gaussian 16 software.[48] In the previous quantum chemical calculations[31] using B3LYP DFT functional[49] and the 6-311++G basis set, the total hyperpolarizability $\beta_{tot}$ values at $\lambda = \infty$ ($\omega = 0$) were calculated to be $1321 \times 10^{-40}$ ($1070 \times 10^{-40}$) m$^4$V$^{-1}$ for a PNPA (NDPA) molecule. The re-examination in this study reproduced almost the same results [$1406 \times 10^{-40}$ ($1057 \times 10^{-40}$) m$^4$V$^{-1}$] using the B3LYP/6-311++G(d,p). However, these calculations using the B3LYP functional do not include the long-range electron interaction properly and tend to overestimate the magnitude of $\beta$ with expanding the π-conjugated system as in the present case of PNPA and NDPA. Instead, recent theoretical simulations using the proper functionals demonstrate the improved agreements with experimental $\beta$ values.[50,51] Thus, we recalculated the $\beta$ using the long-range corrected functional such as the CAM-B3LYP/6-311++G(d,p)[52] as recommended. The revised $\beta_{tot}$ (-$\omega$; $\omega$,0) values of PNPA (NDPA) are $416.8 \times 10^{-40}$ ($521.2 \times 10^{-40}$) m$^4$V$^{-1}$ at $\lambda = \infty$ ($\omega = 0$), $702.7 \times 10^{-40}$ ($862.8 \times 10^{-40}$) m$^4$V$^{-1}$ at $\lambda = 810$ nm, and $808.4 \times 10^{-40}$ ($989.0 \times 10^{-40}$) m$^4$V$^{-1}$ at $\lambda = 730$ nm. It should be also noted that the order of magnitude in the two isomorphous materials is reversed to NDPA > PNPA. This relationship is verified by the EO performance (see Section 2.3), suggesting the importance of the long-range corrections in the cases of PNPA and NDPA. For other organic materials listed in Fig. 1, we calculated the $\beta_{tot}$ (-$\omega$; $\omega$,0) values using the similar functional of CAM-B3LYP/6-311++G(d,p).

The $\beta$ values listed in Table 1 are given in SI units, and the conversion between cgs and SI units is $1 \times 10^{-30}$ esu = $4.18 \times 10^{-40}$ m$^4$V$^{-1}$.

**Birefringence FFMI measurement**

We employed the birefringence FFMI technique to evaluate the EO performance (see Fig. S6 for the detailed setup[35]). When the polar axis of the EO film is set to the direction tilted 45° from the polarizer in the crossed-Nicols configuration, the intensity of transmitted light $I_\perp$ is expressed as

$$I_\perp = I_{\perp 0} + I_{\perp 1} \sin^2\left(\frac{\phi_0}{2}\right), \tag{1}$$

where $\phi_0$ is the phase retardation between the components of elliptically polarized outgoing light along the directions parallel and perpendicular to the polarizer. It is expressed as $\emptyset_0 = \frac{2\pi d}{\lambda} n_{\text{aniso}}$ ($n_{\text{aniso}} = n_3 - n_1$) with the refractive index along the direction parallel (perpendicular) to the polar axis within the (010) plane, $n_{3(1)}$. The point group of PNPA and NDPA is $m$,[31–33] for which the related EO tensors are $r_{33}$, $r_{13}$, and $r_{53}$ when the electric field along the polar axis (|| 3) $E_3$ is applied. The terms of $r_{13}$ and $r_{53}$ are expected to be negligibly small relative to $r_{33}$ because the present systems are highly one-dimensional. Thus, we assume that the change in the anisotropy of the refractive indexes through the EO effect is described as $\Delta n_{\text{aniso}} = -\frac{1}{2} n_3^3 r_{33} E_3$. This leads to the modulation of the outgoing light intensity, $\Delta I_\perp$, as follows,

$$\Delta I_\perp = I_{\perp 1} \left[ \sin^2\left(\frac{\emptyset_0}{2} + \frac{\pi d}{\lambda} \Delta n_{\text{aniso}}\right) - \sin^2\left(\frac{\emptyset_0}{2} - \frac{\pi d}{\lambda} \Delta n_{\text{aniso}}\right) \right]. \tag{2}$$

To sensitively detect the modulation signals, we used the variable retarder as described in Section 2.3. In this case, $\emptyset_0$ in Equations (1)–(2) is replaced with $\emptyset_0 + \emptyset_{\text{ext}}$, where $\emptyset_{\text{ext}}$ is the controllable phase retardation ($0 \leqq \emptyset_{\text{ext}} < 2\pi$) derived from the birefringence of the variable retarder. The slow axis of the variable retarder is parallel to the polar axis of the EO thin film. By fitting of Equations (1) and (2) to $I_\perp$ and $\Delta I_\perp / I_\perp$ measured under the various $\emptyset_{\text{ext}}$ values, $\Delta n_{\text{aniso}}$ ($= -\frac{1}{2} n_3^3 r_{33} E_3$) can be obtained.

In the present study, the applied ac voltage was ±2 V with the modulation frequency of 45 Hz. The gap between the electrodes was 5 μm. The thin-film thickness of PNPA (NDPA) used in the birefringence FFMI measurement was 130 nm (1.2 μm), which was determined by the atomic force microscopy (AFM).

**Acknowledgements**
The authors thank A. Otomo for fruitful discussion. This work was supported by the JSPS Grants-in-Aid for Scientific Research (Grant Nos. 24K17021, 25H00836, and 25H01405). Part of the computation was performed using Research Center for Computational Science, Okazaki, Japan (Project: 24-IMS-C321).

**Author Contributions**
K.S. performed investigation, analysis, and funding acquisition, and wrote the original draft. S.H. performed conceptualization, investigation, analysis, review and editing. Y.S. performed investigation, analysis, review and editing. N.F. performed investigation. T.H. performed calculations for molecules, funding acquisition, review and editing. Y.A. provided resources and performed review and editing. S.I. performed preliminary calculations for crystals, review and editing. J.T. performed conceptualization, supervision, analysis, funding acquisition, review and editing. All authors have read and agreed to the final version of the manuscript.

**References**

1  M. Abarkan, J. P. Salvestrini, M. D. Fontana and M. Aillerie, *Appl. Phys. B*, 2003, **76**, 765–769.
2  N. Margalit, C. Xiang, S. M. Bowers, A. Bjorlin, R. Blum and J. E. Bowers, *Appl. Phys. Lett.*, 2021, **118**, 220501.
3  S. Shekhar, W. Bogaerts, L. Chrostowski, J. E. Bowers, M. Hochberg, R. Soref and B. J. Shastri, *Nat. Commun.*, 2024, **15**, 751.
4  S. Y. Siew, B. Li, F. Gao, H. Y. Zheng, W. Zhang, P. Guo, S. W. Xie, A. Song, B. Dong, L. W. Luo, C. Li, X. Luo and G. Q. Lo, *J. Light. Technol.*, 2021, **39**, 4374–4389.
5  A. Rahim, T. Spuesens, R. Baets and W. Bogaerts, *Proc. IEEE*, 2018, **106**, 2313–2330.
6  G. T. Reed, G. Z. Mashanovich, F. Y. Gardes, M. Nedeljkovic, Y. Hu, D. J. Thomson, K. Li, P. R. Wilson, S. W. Chen and S. S. Hsu, *Nanophotonics*, 2014, **3**, 229–245.
7  A. Liu, L. Liao, D. Rubin, H. Nguyen, B. Ciftcioglu, Y. Chetrit, N. Izhaky and M. Paniccia, *Opt. Express*, 2007, **15**, 660.
8  M. He, M. Xu, Y. Ren, J. Jian, Z. Ruan, Y. Xu, S. Gao, S. Sun, X. Wen, L. Zhou, L. Liu, C. Guo, H. Chen, S. Yu, L. Liu and X. Cai, *Nat. Photonics*, 2019, **13**, 359–364.
9  I. Taghavi, M. Moridsadat, A. Tofini, S. Raza, N. A. F. Jaeger, L. Chrostowski, B. J. Shastri and S. Shekhar, *Nanophotonics*, 2022, **11**, 3855–3871.
10  C. Wang, M. Zhang, X. Chen, M. Bertrand, A. Shams-Ansari, S. Chandrasekhar, P. Winzer and M. Lončar, *Nature*, 2018, **562**, 101–104.
11  Y. Shi, C. Zhang, H. Zhang, J. H. Bechtel, L. R. Dalton, B. H. Robinson and W. H. Steier, *Science (80-. ).*, 2000, **288**, 119–122.
12  L. R. Dalton, P. A. Sullivan and D. H. Bale, *Chem. Rev.*, 2010, **110**, 25–55.
13  S. I. Inoue and A. Otomo, *Appl. Phys. Lett.*, 2013, **103**, 171101.
14  H. Xu, D. L. Elder, L. E. Johnson, Y. de Coene, S. R. Hammond, W. Vander Ghinst, K. Clays, L. R. Dalton and B. H. Robinson, *Adv. Mater.*, 2021, **33**, 2104174.
15  D. L. Elder and L. R. Dalton, *Ind. Eng. Chem. Res.*, 2022, **61**, 1207–1231.
16  V. R. Almeida, Q. Xu, C. A. Barrios and M. Lipson, *Opt. Lett.*, 2004, **29**, 1209.
17  H. Xu, F. Liu, D. L. Elder, L. E. Johnson, Y. De Coene, K. Clays, B. H. Robinson and L. R. Dalton, *Chem. Mater.*, 2020, **32**, 1408–1421.

18  G. W. Lu, J. Hong, F. Qiu, A. M. Spring, T. Kashino, J. Oshima, M. Ozawa, H. Nawata and S. Yokoyama, *Nat. Commun.*, 2020, **11**, 4224.

19  H. Nakanishi, H. Matsuda, S. Okada and M. Kato, *Proc. MRS Int. Meet. Adv. Mater.*, 1989, **1**, 97–104.

20  S. R. Marder, J. W. Perry and W. P. Schaefer, *Science (80-. ).*, 1989, **245**, 626.

21  F. Pan, G. Knöpfle, C. Bosshard, S. Follonier, R. Spreiter, M. S. Wong and P. Günter, *Appl. Phys. Lett.*, 1996, **69**, 13–15.

22  O. P. Kwon, S. J. Kwon, M. Jazbinsek, F. D. J. Brunner, J. I. Seo, C. Hunziker, A. Schneider, H. Yun, Y. S. Lee and P. Günter, *Adv. Funct. Mater.*, 2008, **18**, 3242–3250.

23  M. Jazbinsek, U. Puc, A. Abina and A. Zidansek, *Appl. Sci.*, 2019, **9**, 882.

24  D. Korn, M. Jazbinsek, R. Palmer, M. Baier, L. Alloatti, H. Yu, W. Bogaerts, G. Lepage, P. Verheyen, P. Absil, P. Guenter, C. Koos, W. Freude and J. Leuthold, *IEEE Photonics J.*, 2014, **6**, 2700109.

25  D. R. Kanis, M. A. Ratner and T. J. Marks, *Chem. Rev.*, 1994, **94**, 195–242.

26  J. Liu, C. Ouyang, F. Huo, W. He and A. Cao, *Dye. Pigment.*, 2020, **181**, 108509.

27  G. Clydesdale, R. Docherty and K. J. Roberts, *Comput. Phys. Commun.*, 1991, **64**, 311–328.

28  G. Clydesdale, K. J. Roberts and R. Docherty, *J. Cryst. Growth*, 1996, **166**, 78–83.

29  A. Gavezzotti, *Acc. Chem. Res.*, 1994, **27**, 309–314.

30  A. Gavezzotti and G. Filippini, *J. Phys. Chem.*, 1994, **98**, 4831–4837.

31  G. A. Valdivia-Berroeta, Z. B. Zaccardi, S. K. F. Pettit, S. H. Ho, B. W. Palmer, M. J. Lutz, C. Rader, B. P. Hunter, N. K. Green, C. Barlow, C. Z. Wayment, D. J. Ludlow, P. Petersen, S. J. Smith, D. J. Michaelis and J. A. Johnson, *Adv. Mater.*, 2022, **34**, 2107900.

32  T. E. Souza, A. O. L. Iara Maria Landre Rosa, D. Paschoal, L. J. Q. Maia, H. F. Dos Santos, F. T. Matins and A. C. Doriguetto, *Acta Crystallogr. Sect. B Struct. Sci. Cryst. Eng. Mater.*, 2015, **71**, 416–426.

33  O. S. Bushuyev, T. A. Singleton and C. J. Barrett, *Adv. Mater.*, 2013, **25**, 1796–1800.

34  C. Rader, Z. B. Zaccardi, S. H. E. Ho, K. G. Harrell, P. K. Petersen, M. F. Nielson, H. Stephan, N. K. Green, D. J. H. Ludlow, M. J. Lutz, S. J. Smith, D. J. Michaelis and J. A. Johnson, *ACS Photonics*, 2022, **9**, 3720–3726.

35  K. Sunami, S. Horiuchi, S. Ishibashi and J. Tsutsumi, *Adv. Electron. Mater.*, 2024, **11**, 2400346.

36  H. Sato, H. Miura, F. Qiu, A. M. Spring, T. Kashino, T. Kikuchi, M. Ozawa, H. Nawata, K. Odoi and S. Yokoyama, *Opt. Express*, 2017, **25**, 768.

37  A. Schwarzenberger, A. Mertens, H. Kholeif, A. Kotz, C. Eschenbaum, L. E. Johnson, D. L. Elder, S. R. Hammond, K. O'Malley, L. Dalton, S. Randel, W. Freude and C. Koos,

*49th Eur. Conf. Opt. Commun. (ECOC 2023)*, 2023, **2023**, 859–862.

38 Telcordia, *GR-468-CORE*, 2004.

39 W. Freude, A. Kotz, H. Kholeif, A. Schwarzenberger, A. Kuzmin, C. Eschenbaum, A. Mertens, S. Sarwar, P. Erk, S. Brase and C. Koos, *IEEE J. Sel. Top. Quantum Electron.*, 2024, **30**, 3400222.

40 K. M. Holland, A. Alejandro, D. J. H. Ludlow, P. K. Petersen, M. A. Wright, C. C. Chartrand, D. J. Michaelis, J. A. Johnson and J. E. Patterson, *Opt. Lett.*, 2023, **48**, 5855.

41 M. F. Zaini, I. A. Razak, W. M. Khairul and S. Arshad, *Acta Crystallogr. Sect. E Crystallogr. Commun.*, 2020, **76**, 387–391.

42 S. Makita, A. Saito, M. Hayashi, S. Yamada, K. Yoda, J. Otsuki, T. Takido and M. Seno, *Bull. Chem. Soc. Jpn*, 2000, **73**, 1525–1533.

43 H. Hashimoto, Y. Okada, H. Fujimura, M. Morioka, O. Sugihara, N. Okamoto and R. Matsushima, *Jpn. J. Appl. Phys.*, 1997, **36**, 6754–6760.

44 X. Zhang, X. Jiang, Y. Li, Z. Lin, G. Zhang and Y. Wu, *CrystEngComm*, 2015, **17**, 1050–1055.

45 O. Vakuliuk, B. Koszarna and D. T. Gryko, *Adv. Synth. Catal.*, 2011, **353**, 925–930.

46 B. Zuo, J. Chen, M. Liu, J. Ding, H. Wu and W. Su, *J. Chem. Res.*, 2009, **2009**, 14–16.

47 M. Lemaire, A. Guy, P. Boutin and J. P. Guette, *Synthesis (Stuttg).*, 1989, **10**, 761–763.

48 Gaussian 16, Revision C.02, M. J. Frisch, G. W. Trucks, H. B. Schlegel, G. E. Scuseria, M. A. Robb, J. R. Cheeseman, G. Scalmani, V. Barone, G. A. Petersson, H. Nakatsuji, X. Li, M. Caricato, A. V. Marenich, J. Bloino, B. G. Janesko, R. Gomperts, B. Mennucci, H. P. Hratchian, J. V. Ortiz, A. F. Izmaylov, J. L. Sonnenberg, D. Williams-Young, F. Ding, F. Lipparini, F. Egidi, J. Goings, B. Peng, A. Petrone, T. Henderson, D. Ranasinghe, V. G. Zakrzewski, J. Gao, N. Rega, G. Zheng, W. Liang, M. Hada, M. Ehara, K. Toyota, R. Fukuda, J. Hasegawa, M. Ishida, T. Nakajima, Y. Honda, O. Kitao, H. Nakai, T. Vreven, K. Throssell, J. A. Montgomery, Jr., J. E. Peralta, F. Ogliaro, M. J. Bearpark, J. J. Heyd, E. N. Brothers, K. N. Kudin, V. N. Staroverov, T. A. Keith, R. Kobayashi, J. Normand, K. Raghavachari, A. P. Rendell, J. C. Burant, S. S. Iyengar, J. Tomasi, M. Cossi, J. M. Millam, M. Klene, C. Adamo, R. Cammi, J. W. Ochterski, R. L. Martin, K. Morokuma, O. Farkas, J. B. Foresman and D. J. Fox, *Gaussian, Inc., Wallingford CT,* 2019.

49 A. D. Becke, *J. Chem. Phys.*, 1993, **98**, 5648–5652.

50 L. E. Johnson, L. R. Dalton and B. H. Robinson, *Acc. Chem. Res.*, 2014, **47**, 3258–3265.

51 F. A. Santos, C. E. R. Cardoso, J. J. Rodrigues, L. De Boni and L. M. G. Abegão, *Photonics*, 2023, **10**, 545.

52 T. Yanai, D. P. Tew and N. C. Handy, *Chem. Phys. Lett.*, 2004, **393**, 51–57.

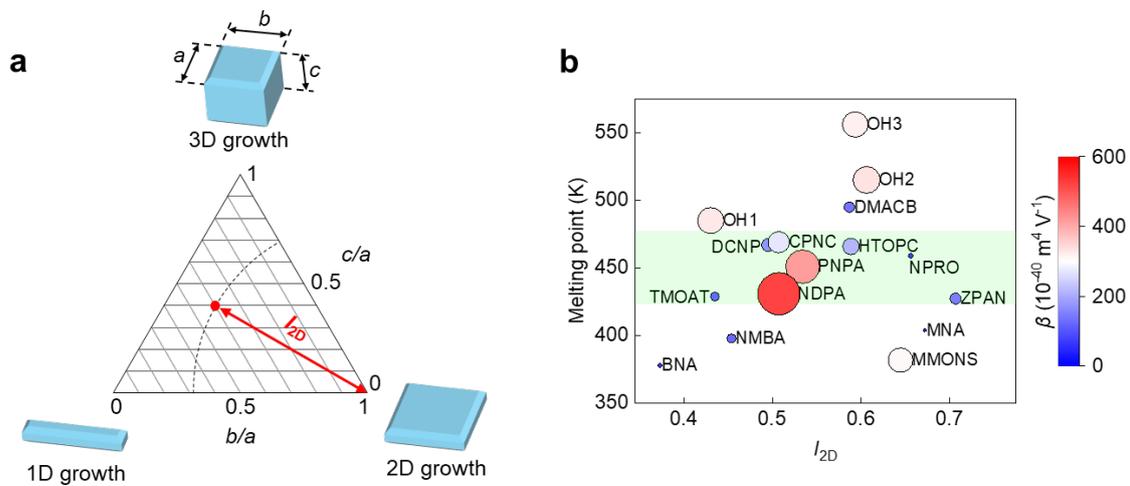

**Figure 1. Screening by crystal habit. a**. Definition of the crystal-habit parameter $I_{2D}$ in a triangle graph as a function of growth anisotropies (see text). $I_{2D}$ corresponds to the distance from the 2D crystal growth in the triangle graph. **b**. Plot of the melting point $T_m$ of EO molecular crystals as a function of $I_{2D}$. The size and color gradation of the symbols represent the magnitude of total hyperpolarizability (non-resonant) $\beta_{tot}$ (see text). The green shaded zone indicates the temperature range suitable for the optoelectronic integration by melt casting (see text).

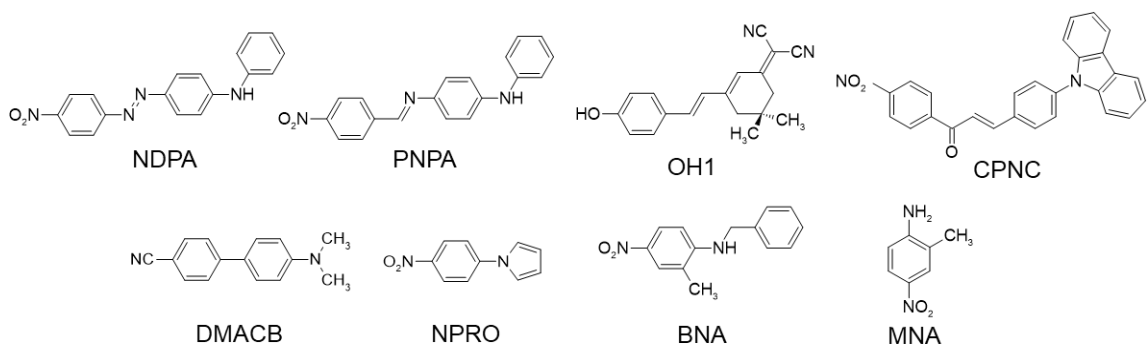

**Figure 2. EO molecular crystals with 2D crystal growth.** Molecular structures of the EO molecular crystals investigated in this study.

| Material | $\beta$ ($10^{-40}$ m$^4$ V$^{-1}$) | $\lambda_{th}$ (nm) | $T_m$ (K) | $T_m - T_s$ (K) |
|---|---|---|---|---|
| NDPA | 521.2 | 665 | 436 | 34 |
| PNPA | 416.7 | 629 | 452 | 75 |
| OH1 | 326.5 | 592 | 482 | >212 |
| CPNC | 270.9 | 565 | 470 | >200 |
| DMACB | 137.5 | 421 | 495 | — |
| NPRO | 74.0 | 434 | 459 | 17 |
| BNA | 64.4 | 495 | 378 | >108 |
| MNA | 50.6 | 498 | 406 | 13 |

**Table 1. Physical properties of EO molecular crystals.** Molecular total hyperpolarizability (non-resonant) $\beta_{tot}$, absorption threshold wavelength $\lambda_{th}$, melting point $T_m$, and the difference between $T_m$ and the solidification point $T_s$ in the EO molecular crystals. $\beta_{tot}$ was calculated by Gaussian software. $\lambda_{th}$ was evaluated by the diffuse reflectance spectroscopy. $T_m$ and $T_s$ were determined by the DSC measurements (see Section 4). For DMACB, the value of $T_m$ - $T_s$ was undefined due to the structural change after re-solidification (see Supporting Information).

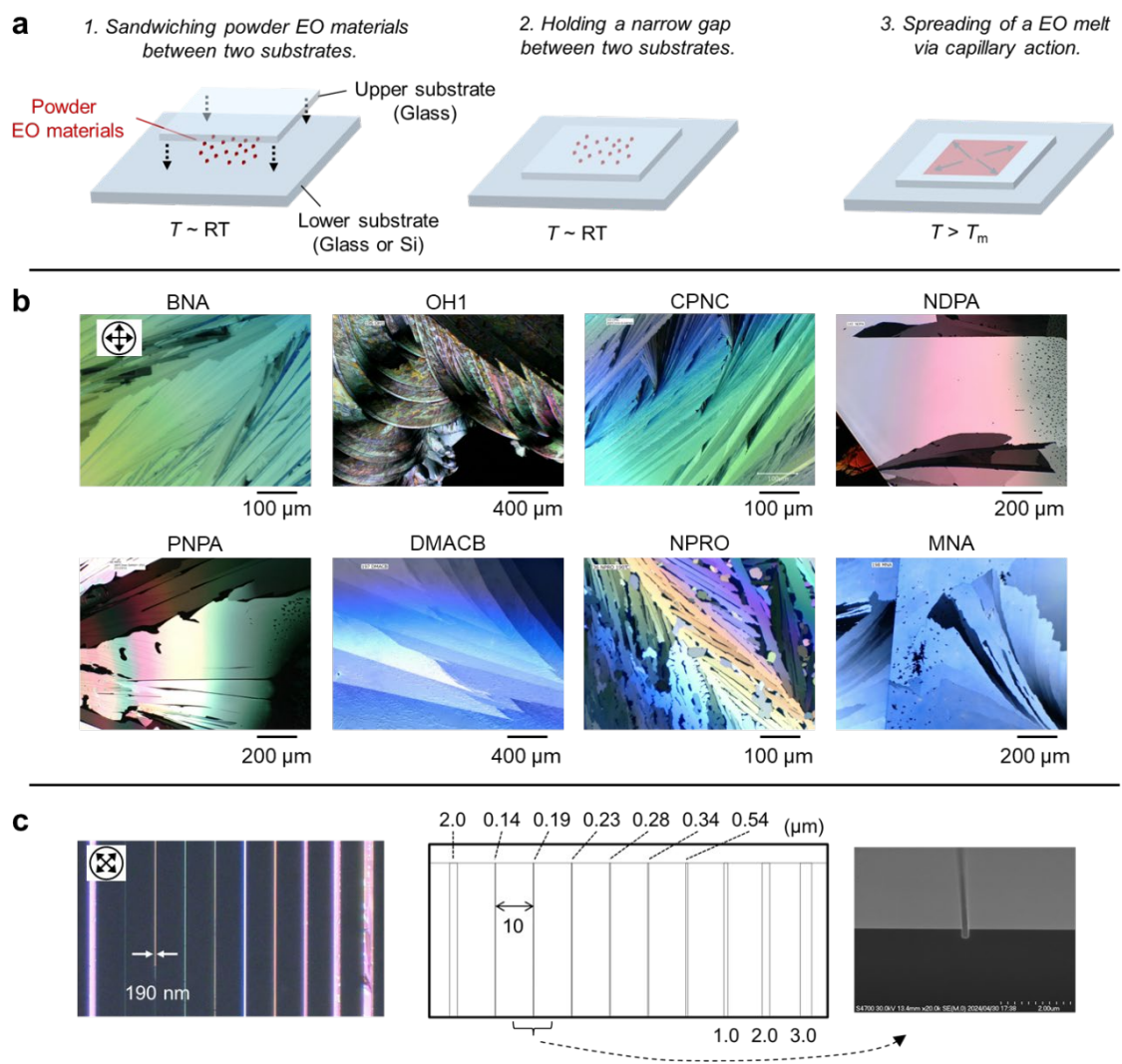

**Figure 3. Thin-film growth by melting casting process. a**. Schematics of thin-film fabrication. **b**. Crossed-Nicols images for thin films of EO molecular crystals fabricated by this process. **c**. PNPA film grown into narrow slots in the silicon substrate. (left) Crossed-Nicols image. (middle) The width and spacing of the slots. (right) The scanning electron microscope (SEM) image of the 190-nm slot.

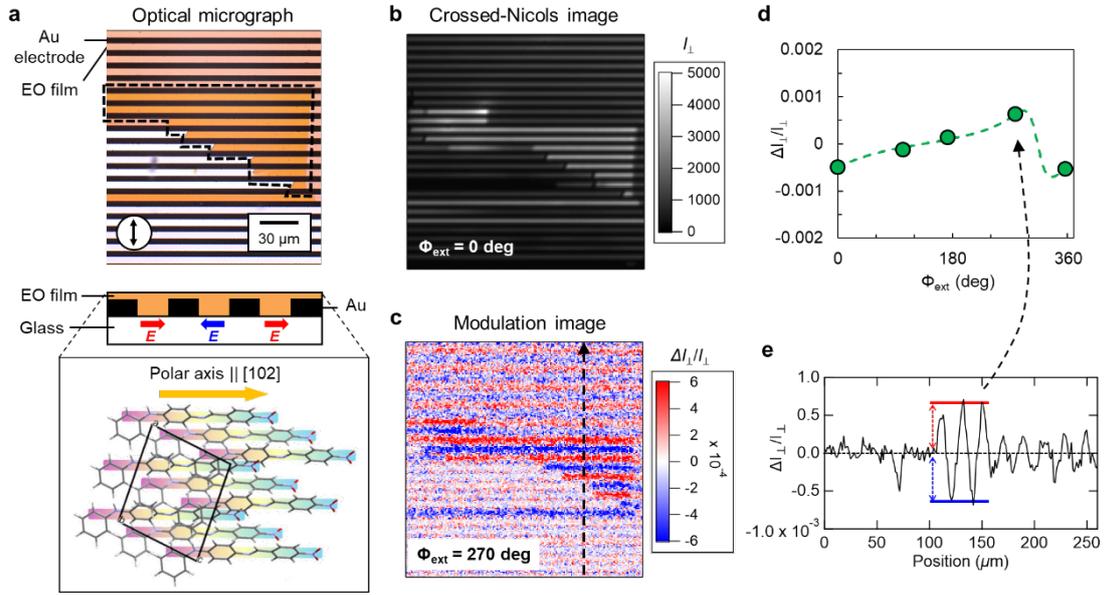

**Figure 4. Evaluation of the electro-optic performance for PNPA thin film at room temperature. a**. Polarized transmission micrograph of PNPA thin film on the glass substrate with interdigitated gold electrodes. The dark regions correspond to the electrodes. The orange-colored region surrounded by the dotted line corresponds to the domain where the polar axis (|| [102]) is parallel to the electric field as shown in the lower panel (Supporting Information). The crystal structure of PNPA shown in the lower panel is viewed along the [010] direction. Crossed-Nicols (**b**) and modulation (**c**) images obtained by the FFMI measurements at $\lambda = 810$ nm under $\varphi_{ext} = 0$ and 270 degrees, respectively. $\varphi_{ext}$ is the phase retardation derived from the variable retarder (see text). **d**. The $\varphi_{ext}$ dependence of the modulation signal $\Delta I_\perp/I_\perp$ at $\lambda = 810$ nm. **e**. Cross-sectional plot of $\Delta I_\perp/I_\perp$ at the dashed arrow in the modulation image of **c**.

| Material | λ (nm) | $n^3r$ (pm V$^{-1}$) |
|---|---|---|
| PNPA | 810 | 690 |
|  | 730 | 1200 |
| NDPA | 810 | 1050 |
|  | 730 | 1250 |

**Table 2. Electro-optic performance in PNPA and NDPA thin films.** The figure of merits for EO performance, $n_3^3 r_{33}$, in PNPA and NDPA thin films evaluated by the FFMI measurements at room temperature.

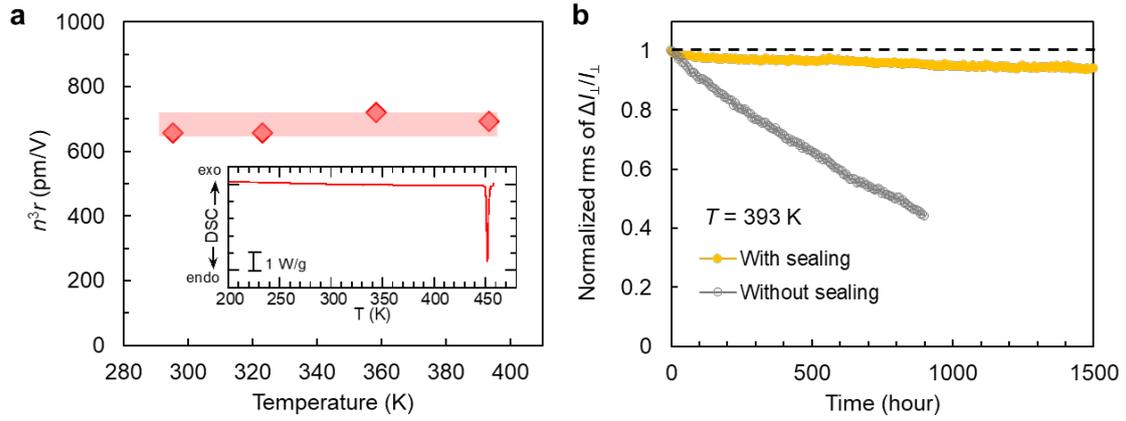

**Figure 5. Thermal stability for electro-optic performance in PNPA thin film. a**. Temperature dependence of the figure of merits for EO performance, $n_3^3 r_{33}$, at $\lambda = 810$ nm. The inset shows the results of the DSC measurements (see Supporting Information). **b**. Plot of the root mean square values of the modulation signal $\Delta I_\perp / I_\perp$ over the area of $>250 \times 250$ μm$^2$ in the EO film fabricated on the interdigitated electrode substrate, measured at 393 K in a nitrogen atmosphere with and without sealing by the glass plate at $\lambda = 810$ nm. The plotted values are normalized to the value at $t = 0$ h.

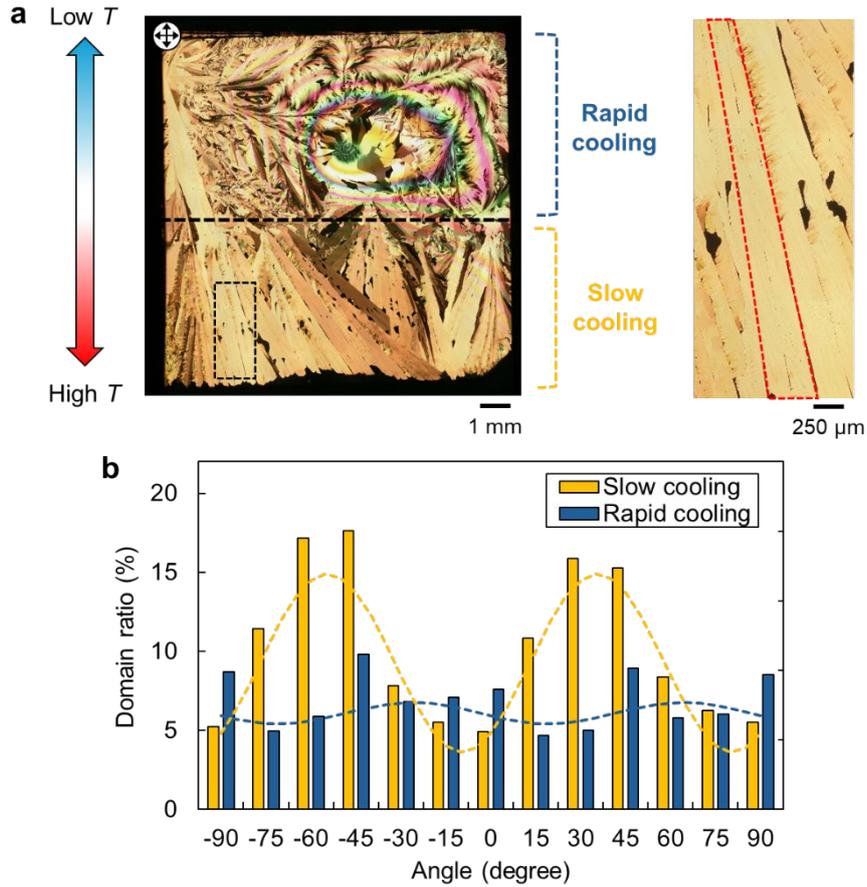

**Figure 6. Unidirectional crystal growth by temperature gradient in PNPA thin film. a**. Crossed-Nicols image of the PNPA film fabricated by the melt casting under the temperature gradient (see text). The dashed black rectangle represents the area shown in the enlarged image on the right. The dashed red box in the enlarged figure is guide for the oriented domain. **b**. Distribution of domain orientation angles relative to the temperature gradient direction (see text).

Supporting Information for

Emerging electro-optic molecular crystals for optoelectronic integration

*Keishi Sunami[1]\*, Sachio Horiuchi[1], Yoriko Sonoda[1], Naomi Fujiki[1], Toshiki Higashino[1], Yuki Atsumi[2], Shoji Ishibashi[1], and Jun'ya Tsutsumi[1]*

[1]*Core Electronics Technology Research Institute (CETRI), National Institute of Advanced Industrial Science and Technology (AIST), Tsukuba, Ibaraki 305-8565 Japan*
[2]*Photonics-Electronics Integration Research Center (PEIRC), National Institute of Advanced Industrial Science and Technology (AIST), Tsukuba, Ibaraki 305-8568 Japan*

\*Corresponding author: k.sunami@aist.go.jp (K.S.), junya.tsutsumi@aist.go.jp (J.T.)

**Supporting Text 1. Second Harmonic Generation (SHG) Measurements.** The Kurtz and Perry powder test [S1] was performed at a fundamental wavelength $\lambda$ = 1.064 μm with the optical path length of 200 μm by measuring the transmitted SHG efficiency relative to urea. The crystals were crushed simply by pushing the wrapping paper sandwiching them and were filtered out to obtain the microcrystalline with particle size of around 63-150 μm. The powdered crystalline croconic acid having the approximate efficiency of 47 times that of urea was repeatedly examined also for regular checking the apparatus with finding the standard deviation of ~30 % on average of 14 specimens.

The observed SHG efficiencies of purified powders relative to urea are NDPA (7.1), PNPA (5.8), OH1 (9.7), CPNC (5.0), DMACB (33, 0.12; before and after melting-re-solidification, respectively), NPRO (110), BNA (250), MNA (700). Note that the NDPA, PNPA, OH1, and CPNC absorb the SH light at $\lambda/2$ = 532 nm, as shown by Fig. S3.

**Supporting Figure 1.** Molecular structures of the molecular crystals listed in Fig. 1b but not shown in Fig. 2.

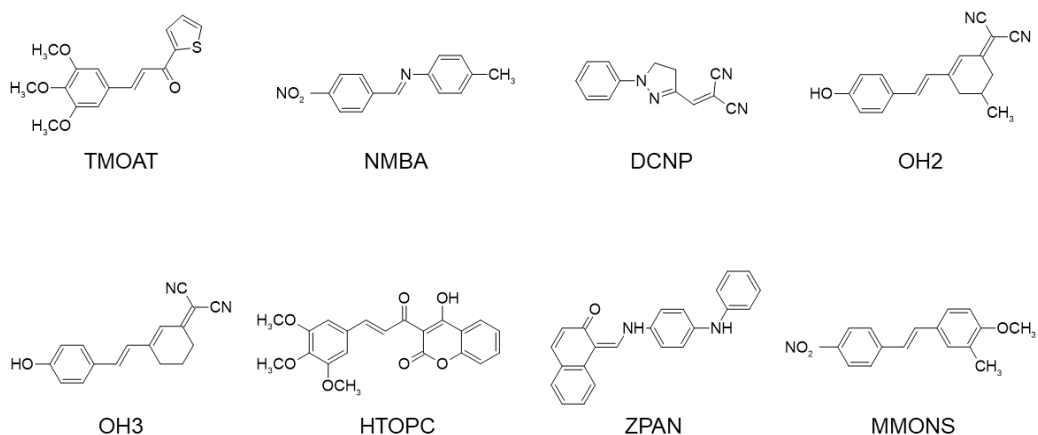

**Supporting Figure 2. Differential scanning calorimetry (DSC) thermographs.** Arrows indicate the directions of temperature changes at a rate of 5 K min$^{-1}$. The inset to panel (f) shows that the phase transition observed before melting disappears below $T_\text{m}$ in the second heating runs, suggesting a formation of unknown polymorph after re-solidification. This observation is consistent with the SHG activity being lost with this treatment.

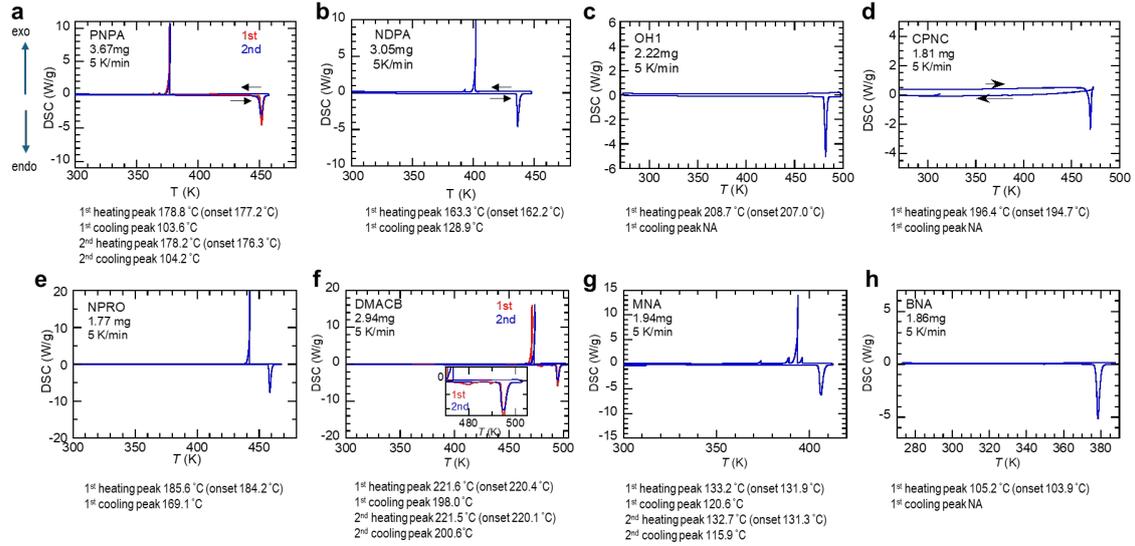

**Supporting Figure 3. Kubelka-Munk (K-M) absorption spectra.**

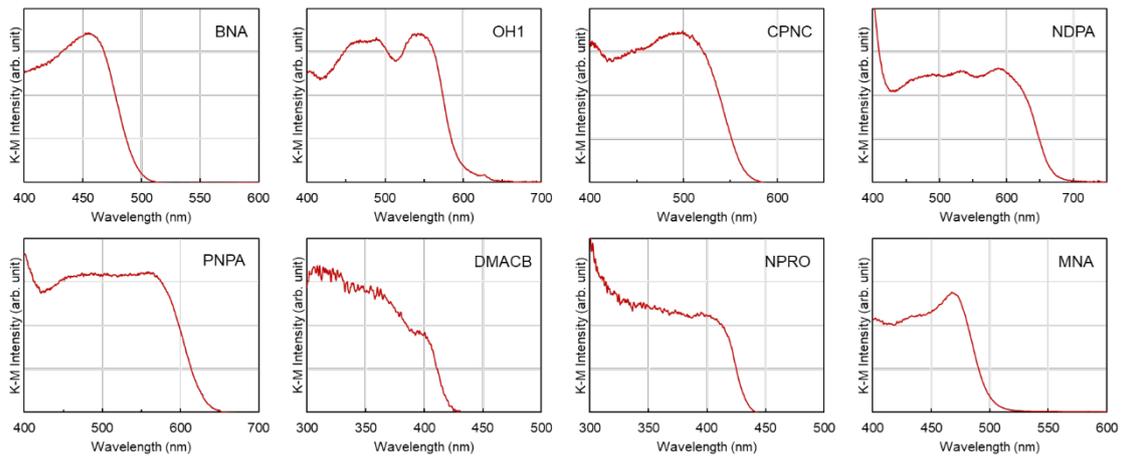

**Supporting Figure 4. Polarized microscope images for PNPA and NDPA thin films.** Crossed-Nicols and polarized microscope images for PNPA (upper panels) and NDPA (lower panels) pictured in the transmission setup. The angle between the polarizer (set before the thin film) and the [001] direction (∥ *c*) is described at the bottom.

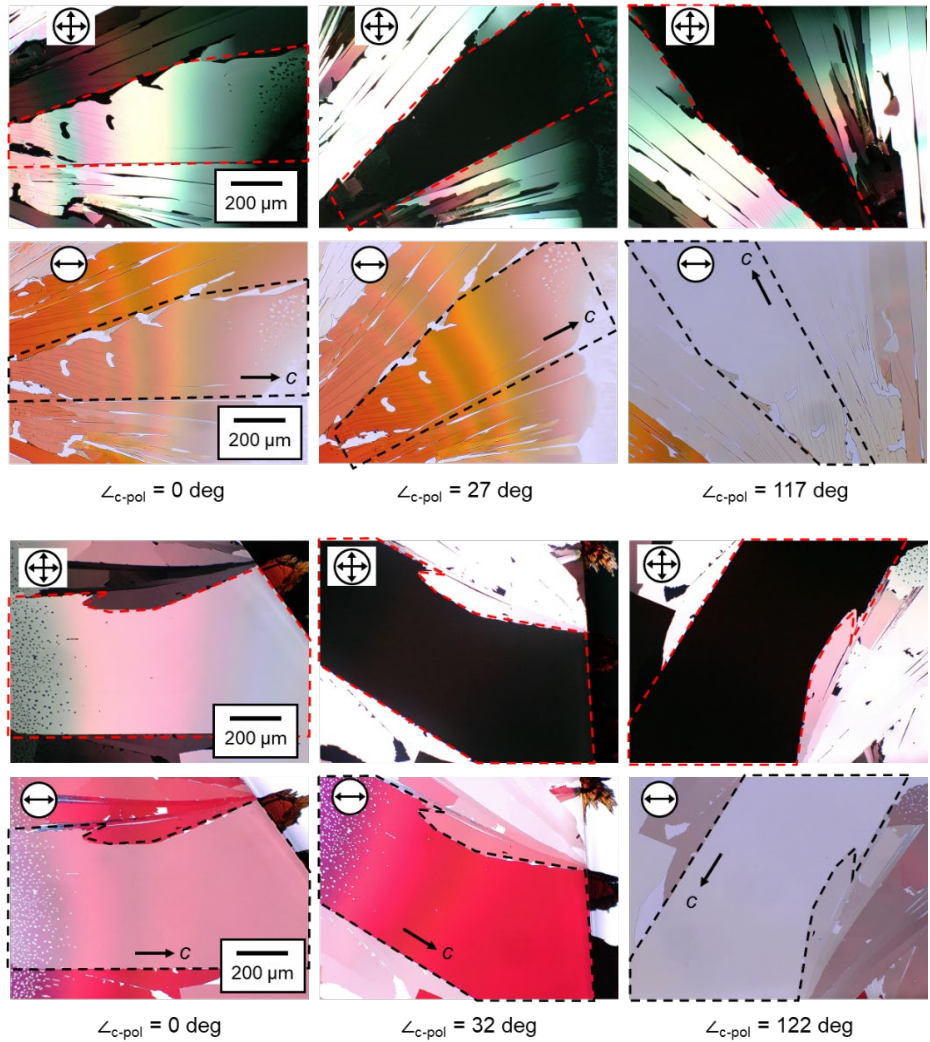

**Supporting Text 2. Determination of the direction of polarity in EO thin films.** According to the crystal structure analysis, the polar axis of PNPA is parallel to the [102] direction, which is inclined by about 30 degrees relative to the [001] direction (corresponding the crystal growth axis). Because a strong absorption band due to intramolecular transition at approximately 500 nm exists along the polar axis in PNPA (Table 1), the polarized microscope image appears to be orange-red colored when the polarizer is parallel to the polar axis of EO films. Additionally, the crossed-Nicols image becomes completely dark at a similar angle because the polar axis is expected to be the dielectric principal axis. The obtained PNPA single crystalline film by the melt casting exhibits these features when the angle between the polarizer and the [001] axis is about 27 degrees (surrounded by the broken line in Fig. S4), indicating the exposure surface plane of the (010) plane with the in-plane polar axis. For NDPA, we observed the similar behaviors at the angle between the polarizer and the [001] axis of 32 degrees, also indicative of the (010) surface film.

**Supporting Figure 5. Filling into narrow slots for NDPA thin film.** Crossed-Nicols image of the NDPA films into the narrow slots in the silicon substrate pictured in the reflection setup.

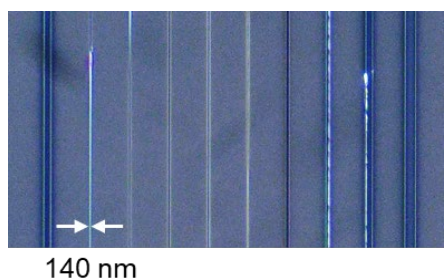

140 nm

**Supporting Figure 6. Setup for modulation imaging.** The EO thin film is positioned such that the polar axis is oriented at 45 degrees relative to the polarizer in the crossed-Nicols configuration.

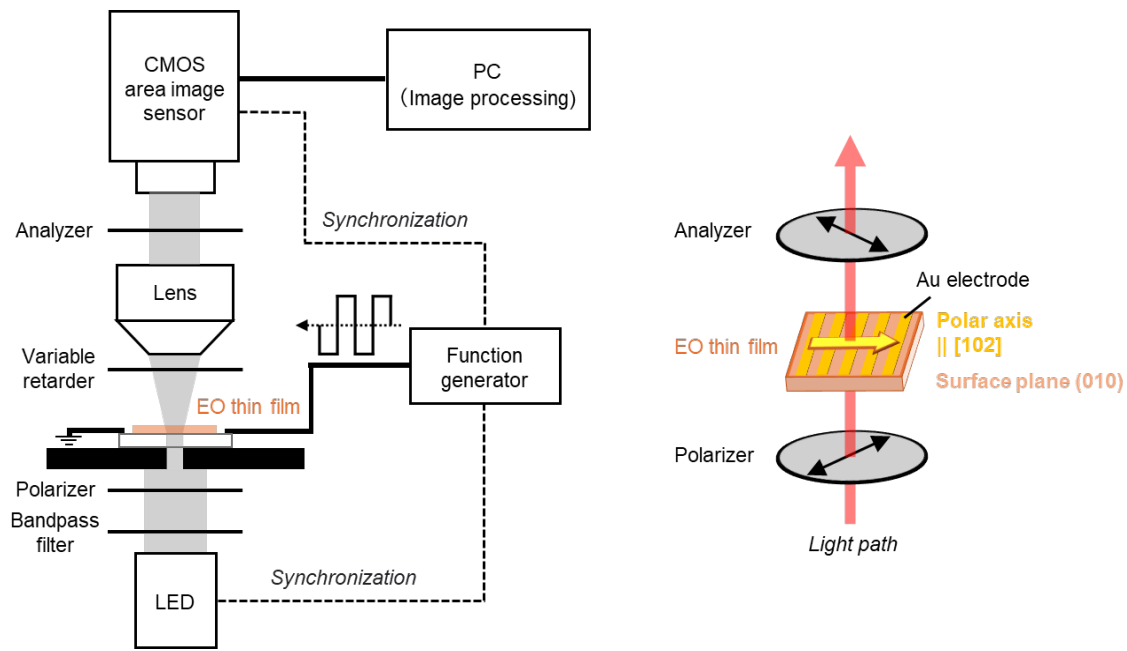

**Supporting Figure 7. Long-term thermal stability tests of PNPA thin film with and without sealing.** Crossed-Nicols ($I_\perp$) and modulation ($\Delta I_\perp/I_\perp$) images for PNPA thin film fabricated on the interdigitated electrode substrate, obtained by the FFMI measurements at $\lambda = 810$ nm (the displayed images represent a part of the entire measurement area). For the sealed condition, the film was covered with a glass plate. The FFMI measurements were conducted under the phase retardation derived from the variable retarder $\varphi_\text{ext}$ of ~180 and ~90 degrees for the sealed and unsealed measurements, respectively. These values were selected to clearly detect the $\Delta I_\perp/I_\perp$ signals in each setup. The $\varphi_\text{ext}$ values that maximize $I_\perp$ and $\Delta I_\perp/I_\perp$ differ by ~$\pi$,[S2] leading to a dark $I_\perp$ image under the condition optimized for observing the $\Delta I_\perp/I_\perp$ signal.

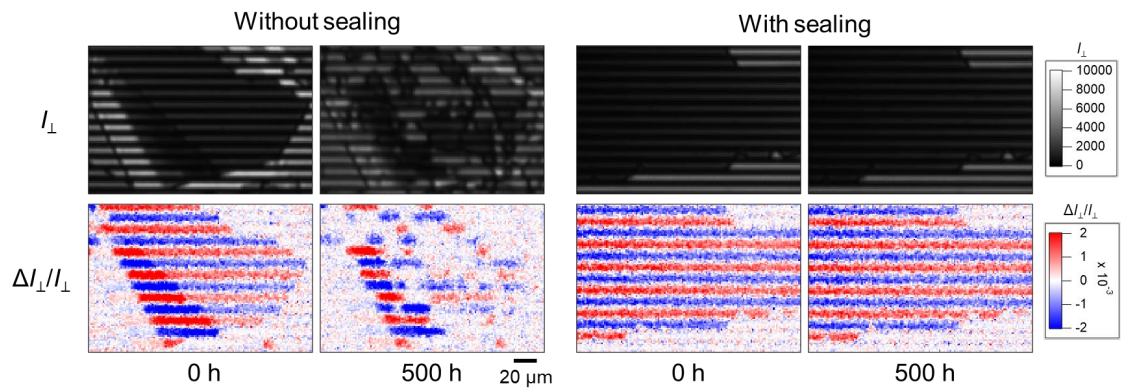

**Supporting references**

[S1] S. K. Kurtz and T. T. Perry, *J. Appl. Phys.* **1968**, *39*, 3798.
[S2] K. Sunami, S. Horiuchi, S. Ishibashi, J. Tsutsumi, *Adv. Electron. Mater.* **2024**, *11*, 2400346.